\documentstyle[12pt,amssymb]{article}

\newcommand{\beq}[1]{\begin{equation}\label{#1}}
\newcommand{\eeq}{\end{equation}}
\newcommand{\bear}[1]{\begin{eqnarray}\label{#1}}
\newcommand{\ear}{\end{eqnarray}}

\newcommand{\rf}[1]{(\ref{#1})}
\newcommand{\nl}{ {\hfill \break} }
\newcommand{\np}{ {\newpage } }

\newcommand{\Iff}{ {\Leftrightarrow } }

\newcommand{\partl}{ {\partial } }
\newcommand{\cl}{ { \mathrm{cl} }  }
\newcommand{\clc}{ { \mathrm{cl}_c} }

\newcommand{\N}{ \mbox{\rm I$\!$N} }
\newcommand{\R}{ \mbox{\rm I$\!$R} }
\newcommand{\Z}{ \mbox{$\mathbb Z$} }
\newcommand{\Diff}{ \mbox{\rm Diff} }

\newcommand{\SO}{ \mbox{\rm SO} }

\newcommand{\Imp}{ \mbox{ $\Rightarrow$ } }

\newcommand{\Int}{ {\mathrm int} }
\newcommand{\Img}{ {\mathrm im} }
\newcommand{\C}{ {\mathcal{C}} }
\newcommand{\Cr}{  {\mathcal{C}}^r }
\newcommand{\CO}{ {\mathcal{C}}^0 }
\newcommand{\CI}{ {\mathcal{C}}^1 }
\newcommand{\Cinf}{ {\mathcal{C}}^\infty }

\renewcommand{\setminus}{-}

\def\be{\begin{equation}}
\def\ee{\end{equation}}
\def\bea{\begin{eqnarray}}
\def\eea{\end{eqnarray}}

\DeclareFontFamily{U}{rsfs}{}         
\DeclareFontShape{U}{rsfs}{m}{n}{<5> rsfs5 <6><7> rsfs7          %
  <8><9><10><10.95><12><14.4><17.28><20.74><24.88> rsfs10}{}     %
\DeclareMathAlphabet{\mathfs}{U}{rsfs}{m}{n}                     %
\newcommand{\mfs}[1]{\mathfs {#1}}                               %

\newcommand{\sC}{ {\mfs C}}
\newcommand{\sK}{ {\mfs K}}
\newcommand{\sF}{ {\mfs F}}
\newcommand{\sP}{ {\mfs P}}
\newcommand{\sT}{ {\mfs T}}
\newcommand{\sE}{ {\mfs E}}

\newcommand{\olcirc}[1]{\vbox{\mathsurround=0pt                %
  \skip1=0pt plus 1 fil  \skip3=0pt plus 3fil                    %
  \ialign{##\crcr$\hskip\skip3\scriptstyle\circ\hskip\skip1$\crcr%
    \noalign{\kern1pt\nointerlineskip}$\displaystyle{#1}$\crcr}} %
  \vphantom{#1}}                                                 %
\newcommand{\oscirc}[1]{\vbox{\mathsurround=0pt                %
  \skip1=0pt plus 2 fil  \skip3=0pt plus 2fil                    %
  \ialign{##\crcr$\hskip\skip3\scriptstyle\circ\hskip\skip1$\crcr%
    \noalign{\kern1pt\nointerlineskip}$\displaystyle{#1}$\crcr}} %
  \vphantom{#1}}                                                 %
\newcommand{\dcup}{ {\oscirc \cup}}

\begin{document}
\title{\bf\Large Cones and causal structures \\
on topological and differentiable manifolds}
\author
{M. Rainer\\
Center for Gravitational Physics and Geometry, 104 Davey Laboratory,\\
The Penn State University, University Park, PA 16802-6300, USA \\
and\\
Mathematische Physik I, Mathematisches Institut,\\
Universit\"at Potsdam, PF 601553, D-14415 Potsdam, Germany
}

\date{March 31, 1999}

\maketitle
\centerline{\bf Abstract}
\vspace{0.5cm}
{
General definitions for causal structures
on manifolds
of dimension $d+1>2$ are presented
for the topological category and for any
differentiable one.

Locally, these are given as cone  structures
via local (pointwise) homeomorphic
or diffeomorphic abstraction from
the standard null cone variety in $\R^{d+1}$.
Weak ($\sC$) and strong ($\sC^m$)
local cone (LC) structures refer to
the cone itself
or a  manifold thickening of the cone respectively.

After introducing cone (C-)causality,
a causal complement with reasonable duality properties
can be defined.
The most common causal concepts of space-times are generalized
to the present topological setting.
A new notion of precausality precludes inner boundaries within
future/past cones.

LC-structures, C-causality,
a topological causal complement, and precausality may be useful
tools in conformal and background independent formulations
of (algebraic) quantum field theory and quantum gravity.
\nl
\nl
\nl
\np

\section{Introduction} 
While classical general relativity usually
employs a Lorentzian space-time metric,
all genuine approaches to quantum gravity
are free of such a metric background.
This poses the question whether there
still exists a notion of  structure which captures
some essential features of light cones and
their mutual relations in manifolds
in a purely topological manner
without a priori recursion to a Lorentzian
metric or a conformal class of such metrics.
Below we will see that the answer is positive.

It is a well known folk theorem that the causal structure
on a Lorentzian manifold determines its
metric up to conformal transformations.
In  \cite{HKM,Mal}
a path topology  for strongly causal space-times was defined
which then determined their differential, causal, and conformal structure.
In \cite{FKN}
it was shown that the conformal class of a
Lorentzian metric can be {\em re}constructed from the
characteristic surfaces of the manifold.
Similarly  \cite{Ehr91} gives a nice proof that the null
cones determine the Lorentzian metric  (modulo global sign)
up to a conformal factor.
All these previous results already indicate
that the notion of a causal structure could exist
indeed in a different and possibly more general setting than
that of Lorentzian space-times.
However all the previously mentioned investigations
in the literature  assume a priori the existence of
some  undetermined Lorentzian metric
and then show that it can be determined modulo conformal
transformation uniquely
by some other structure.

Motivated by the requirements
on suitable structures for a theory
of quantum gravity, in this paper new
notions of causal structure are developed
which do not assume
a priori existence of any (Lorentzian) metric or
conformal metric but rather work on
arbitrary topological and differential manifolds.

In Section II weak ($\sC$) and strong ($\sC^m$) local cone (LC)
structures are defined on any topological (or differentiable)
manifold $M$. These structures are given by
continuous (or differentiable) families of pointwise
homeomorphisms from the standard null cone variety in $\R^{d+1}$
or a manifold thickening thereof respectively into $M$. In the
differentiable case it turns out that a strong LC structure
implies the existence of a conformal Lorentzian metric, while a
weak LC structure already implies its uniqueness should one exist.
However the
metric resulting from a strong LC structure
contains only pointwise information about the asymptotic structure
of the cone at the vertex.
Within a given manifold thickening of the cone at a given point of $M$,
the cone in any neighborhood of the vertex
need a priori not at all be related to the null structure spanned
by the null geodesics of this metric.
However, if such a relationship holds in some region,
then all the cones in that region are consistent with each other
and this way yield a notion of causality.

Section III provides precisely those definitions of causality
which allow to formulate the consistency of different strength
for cones at different points, in some or any open region in $M$.
Cone (C-) causality
allows first of all
the definition of a causal complement with reasonable properties.
It enables us also to define
in a topological (differentiable) manner
spacelike, null, and timelike curves.
We discuss C-causality also  in the particular context
of a fibration.
Generalizations of the most common
causality notions for space-times
in purely topological terms are provided.
In the case of Lorentzian manifolds these notions
 agree with the usual ones and they assume their
usual hierarchy.
Finally, precausality is defined as a notion which
makes the future and the past of any cone homeomorphic to
the future  and the past  of the standard cone
$\sC$ in $\R^{d+1}$ respectively.

The discussion points out some of the major open issues
which require further investigation.
It addresses also the issues of causal diffeomorphisms,
foliations, and possible restrictions of the cone structure and
causality from the manifold to an embedded  graph therein,
giving also
a perspective for possible applications in conformal and
background independent quantum field theories
and quantum gravity.

Here and below a CAT manifold refers to a
Hausdorff ($T_2$) space with CAT  structure,
where CAT$=\CO$ (the topological category) or CAT$\subset\CI$
(any differentiable category).
If CAT$\subset\CI$,  a CAT homeomorphism is a diffeomorphism
and a CAT continuous map is a differentiable map.
For differentiable categories we also define
CAT$_{-1}:=\Cr$ if CAT$=\C^{r+1}$,
CAT$_{-1}:=\C^\infty$ if CAT$=\C^\infty$,
and CAT$_{-1}:=\C^\omega$ if CAT$=\C^\omega$.
\section{Local cone (LC) structures of manifolds} 
\setcounter{equation}{0}
In this section we derive local notions of a cone structure
on a topological $d+1$-dimensional
manifold $M$ (CAT$\subset\CO$).
Let
\beq{scone}
\sC:=\{x\in\R^{d+1}:x_0^2=(x-x_0 e_0)^2\} ,
\sC^+:=\{x\in \sC:x_0\geq 0\} ,
\sC^- :=\{x\in \sC:x_0\leq 0\}
\eeq
be the standard (unbounded double) light cone,
and the forward and backward subcones in $\R^{d+1}$, respectively.

The standard open interior and exterior of $\sC$ is defined as
\beq{scone2}
\sT:=\{x\in\R^{d+1}: x_0^2>(x - x_0 e_0)^2\} ,
\sE:=\{x\in\R^{d+1}: x_0^2<(x - x_0 e_0)^2\} .
\eeq
A \emph{manifold thickening} with thickness $m>0$ is given as
\beq{thickcone}
\sC^m:=\{x\in\R^{d+1}:|x_0^2-(x - x_0 e_0)^2|<m^2\} ,
\eeq
The characteristic topological data of the standard cone
is encoded in the topological relations of all its
manifold subspaces
(which includes in particular also the singular vertex $O$)
and among each other.

Typical (CAT) manifold subspaces of $\sC$ are the standard future and past cones
$\sC^\pm$, and the standard light rays
\beq{ray}
l(n):=\{x\in \sC: x_{0} = (x,n)\} ,
\eeq
where $n\in S^{d-1}\subset p$ is a normal direction
in the $d$-dimensional hyperplane
$p:=\{x\in\R^{d+1}: (x,y)=0\ \forall y\in a \}$
perpendicular
to the cone axis
$a:=\{x\in\R^{d+1}: x=\lambda e_{0},\lambda\in \R\}$.

The topological relations between all the CAT manifold subspaces of the cone
are the natural data which will be required to be conserved
under a homeomorphism of the cone as a topological space
into the manifold $M$ at any point $p$.

Let $\tau$ denote the closed sets of the manifold
topology of $\sC-O$.
The set $\sC$ can either inherit the induced topology $\tau_1$ from $\R^{d+1}$
which is $T_1$ but not $T_2$ (Hausdorff) or it can be equipped
with a more coarse subtopology defined in terms of closed sets as
$\tau_2:=\{\{0\}\cup V: V\in\tau\}\bigcup\{V\in\tau\}$
which is Hausdorff. However $\tau_2$ places
geometrically unnatural restrictions
on possible submanifolds of $\sC$. Hence, unless specified
otherwise, $\sC$ will be equipped with $\tau_1$.

\textbf{Definition 1:}
Let $M$ be a CAT manifold.
A (CAT) \emph{(null) cone} at $p\in \Int M$ is
the homeomorphic image $\sC_p:=\phi_p \sC$
of a homeomorphism of topological spaces
$\phi_p: \sC \to \sC_p \subset  M $
with $\phi_p(0)=p$, such that
\\
(i) every (CAT) submanifold $N\subset \sC$
is mapped (CAT) homeomorphically on
a submanifold  $\phi_p(N)\subset M$,
\\
(ii) for any two submanifolds $N_1,N_2\subset \sC$
there exist homeomorphisms
$ \phi_p(N_1) \cap \phi_p(N_2) \cong N_1 \cap N_2$
and $\phi_p(N_1) \cup \phi_p(N_2)\cong N_1 \cup N_2$
of (CAT) manifolds if these are (CAT) manifolds
and of topological spaces otherwise, and
\\
(iii) if CAT$\subset\CI$ then
for any two CAT curves $c_1,c_2:]-\epsilon,\epsilon[\to\sC$
with $c_1(0)=c_2(0)=p$ it holds
$T_0 c_1=T_0 c_2 \Rightarrow
T_p(\phi_p\circ c_1)|_{ ]-\epsilon,\epsilon[ }
=T_p (\phi_p \circ c_2)|_{ ]-\epsilon,\epsilon[ }$.

Condition (iii) says that in the differentiable case
the well defined notion of transversality of intersections
at the vertex is preserved by $\phi_p$.

On each homeomorphic cone $\sC_p$ at any $p\in \Int M$,
the topology $\tau_1$ or $\tau_2$ of $\sC$ yields under $\phi_p$ likewise
a non-Hausdorff $T_1$   topology $\phi_p(\tau_1)$
or a $T_2$ one $\phi_p(\tau_2)$.
However, $\phi_p\circ\tau_2$ would unnaturally restrict the possible
submanifolds
of $\sC$, while $\phi_p\circ\tau_1$ is consistent with the topology
induced from $M$.

\textbf{Definition 2:}
An \emph{(ultraweak) cone structure} on $M$ is
an assignment
$\Int M\ni p\mapsto\sC_p$ of a cone
at every $p\in\Int M$.

A cone structure on $M$ can in general
be rather wild
with cones at different points totally unrelated
unless we impose a topological connection
between the cones at different points.
Most naturally the connection is provided by
continuity of the family $\{\sC_p\}$.
This allows to define a local cone (LC) structures.

\textbf{Definition 3:} Let $M$ be a CAT manifold.
A \emph{weak ($\sC$) local cone (LC) structure}   on $ M$ is
a  cone structure which is (CAT) continuous
i.e. $\{p\mapsto\sC_p\}$ is a (CAT) continuous family.

Given a cone structure one wants to know first of all
under which conditions, for given $p\in\Int  M$
an exterior and interior of the cone can be distinguished
\emph{locally}, i.e. for any open neighborhood $U\ni p$
within  $(M-\sC_p)\cap U$.

\textbf{Proposition 1:}
Let $\forall p\in\Int  M$ exist
open (CAT) submanifolds
$\sT_p$ and $\sE_p$
such that
the interior of $M$ decomposes in the disjoint union
$\Int M =\sC_p \dcup \sT_p \dcup \sE_p$.
\\
(i)
Then $\sT_p$ and $\sE_p$ can be topologically distinguished
locally in any neighborhood of the vertex $p$ if and only if
for any neighborhood $U \ni p$ it holds
$(\sT_p|_U)\not\cong(\sE_p|_U)$.
\\
(ii)
Given any neighborhood $U \ni p$ assume
$\exists$ $k\in\N_0: \Pi_{k}(\sT_p|_U)\neq \Pi_{k}(\sE_p|_U)$.
Then $\sT_p$ and $\sE_p$ can be topologically distinguished
locally in any neighborhood of the vertex $p$.
\\
\emph{Proof:}
(i) follows from $U-\sC_p|_U=\sT_p|_U \dcup \sE_p|_U$.
(ii) holds because homotopy groups are topological
invariants.
\hfill \mbox{$\square$}

Note that, although $\sC_p=\phi_p(\sC)$, $\sT$ and $\sE$ here need
not be homeomorphic to $\phi_p(\sT)$ and $\phi_p(\sE)$ respectively.
The notion of precausality (see below in Section III) is set up
to ensure $\sE_p\cong \phi_p(\sE)$.

A weak LC structure at each point $p\in\Int  M$
defines
a characteristic topological space $\sC_p$ of codimension $1$
which is Hausdorff everywhere but at $p$.
In particular $\sC_p$ does not contain any
open $U\ni p$ from the manifold topology of $M$.
However stronger structures can be defined as follows.

\textbf{Definition 4:}
Let $M$ be a CAT manifold.
A (CAT) \emph{(manifold) thickened cone} of thickness $m>0$ at $p\in\Int  M$
is the (CAT) homeomorphic image $\sC^{m}_{p}:=\phi_p \sC^m$
of a (CAT) homeomorphism of manifolds
$\phi_p: \sC \to \sC_p \subset  M $
with $\phi_p(0)=p$.

Note that due to the manifold property
the notion of a thickened cone is much more simple than that
of a cone itself.
It also clear that now the only consistent
topology on $\sC\subset\sC_{p}$
is $\tau_1$ and correspondingly
$\phi_p(\tau_1)$ on $\sC_{p}\subset\sC^{m}_{p}$.

\textbf{Definition 5:}
A \emph{thickened cone structure} on $M$ is
an assignment
$\Int M\ni p\mapsto\sC^{m(p)}_{p}$ of a
thickened cone
at every $p\in\Int  M$.

Note that in general the thickness $m$
can vary from point to point in $M$
Here $m:M\to\R_+$ is an a priori not necessarily
continuous function.
However an important case even more special than the continuous one
is that of constant $m$.
\\
\textbf{Definition 6:}
A \emph{homogeneously thickened cone structure} on $M$ is
a thickened cone structure
$\Int M\ni p\mapsto\sC_{p}$ with constant thickness $m$.

Although homogeneity might be too restrictive, at least
continuity of structures on $M$
is a natural  assumption in the topological category.
\\
\textbf{Definition 7:} Let $M$ be a CAT manifold.
A \emph{strong  ($\sC^m$) LC structure}  on $ M$ is
a  (CAT) continuous family
of (CAT) homeomorphisms
$\phi_p: \sC^m \to \sC^{m(p)}_{p} \subset M $
with $\phi_p(0)=p$ and
such that the thickness $m$ is a CAT function on $M$.

In particular the conditions of (ii) in Proposition 1 apply
for all manifolds of dimension $d+1>2$
with a strong LC structure,
while a weak LC structure at $p\in \Int M$
may not be able to distinguish
$\sT_p|_U$ and $\sE_p|_U$
within any $U\ni p$.

\textbf{Theorem 1:}
Let $M$ carry a strong LC structure.
At any $p\in \Int M$ there exists an open $U\ni p$
such that:
\\
For $d:=\dim M-1>0$
it is  $|\Pi_{0}(\sT_p|_U)|=2$ and
$\Pi_{d-1}(\sE_p|_U)=\Pi_{d-1}(S^{d-1})$,
\\
for $d>1$ it is
$\Pi_{d-1}(\sT_p|_U)=0$ and $|\Pi_{0}(\sE_p|_U)|=1$,
\\
for $d=1$ it is
$\Pi_{d-1}(\sT_p|_U)=\Pi_{d-1}(\sE_p|_U)=\Pi_{0}(S^{0})$, i.e.
$|\Pi_{0}(\sT_p|_U)|=|\Pi_{0}(\sE_p|_U)|=2$,
\\
and in dimension $d=0$ it is
$\sT_p=\sE_p=\emptyset$.
\\
\emph{Proof:}
For all $p\in \Int M$ the strong LC structure
provides a thickened cone $\sC^{m(p)}_p$.
Since $m(p)>0$, $\sC^{m(p)}_p$ contains always a neighborhood $U\ni p$
homeomorphic to a neighborhood $\phi^{-1}_p(U)\ni 0$
of the standard cone which in any dimension has the
desired properties.
\hfill \mbox{$\square$}

At any interior
point $p\in\Int  M$
the open exterior $\sE_p$ and
the open interior $\sT_p$ of the cone $\sC_p$ are locally
 topologically distinguishable for $d>1$,
indistinguishable for $d=1$, and empty for $d=0$.
With a strong LC structure
$\sT_p|_U \neq\sE_p|_U \forall U\ni p$ $\iff$ $d+1>2$,
whence locally in any neighborhood $U\ni p$
the interior and exterior of $\sC_p\cap U$ at $p$ in $U$
has an intrinsic invariant meaning.
$\sC_p|_U$ divides $U\setminus \sC_p|_U$ in three
open submanifolds,
a non-contractable exterior $\sE_p|_U$,
plus two contractable connected components of
$\sT_p=:\sF_p|_U\cup\sP_p|_U$, the local future
$\sF_p|_U$ and the local past $\sP_p|_U$ with
$\partl \sF_p|_U=\sC^+_p|_U$ where $\sC^+_p:=(\phi_p \sC^+)$
and $\partl \sP_p|_U=\sC^-_p|_U$ where $\sC^-_p:=\phi_p \sC^-$
respectively.
This rises also the question if and how $\sF_p$
and $\sP_p$ or their local restriction to $U\ni p$
can be distinguished.
This problem is solved by a topological ${\Z}_2$ connection
(see also Section III below).


Given a strong LC structure,
a local (conformal) metric can always be proven to exist
on any differentiable manifold $M$ with CAT$\subset\CI$.
Within such CAT,
any metric $\eta$ on $\R^{d+1}$ can be restricted to
$\sC^m$ and pulled back
pointwise along $(\phi_p)^{-1}$ to a metric $g$
on $\sC^{m(p)}_{p}$.
The CAT continuity of the family $\{p\mapsto\sC^{m(p)}_{p}\}$
implies CAT$_{-1}$ continuity of the
family $\{ p\mapsto g|_{\sC^{m(p)}_{p}} \}$.
So we can extract a CAT$_{-1}$ continuous metric $\{ p\mapsto g_p \}$.

Here we are interested in particular only in Lorentzian metrics
which are \emph{locally compatible} with a (weak or strong) LC structure.
The Minkowski metric $\eta$ is locally compatible with the cone $\sC$
in the sense that $\eta_0(v,v)=0\Leftrightarrow v\in T_0 N$,
with arbitrary submanifold $N\subset\sC\subset\R^{d+1}$ such that  $(0,v)\in TN$.
Correspondingly, a Lorentzian metric $g$ is said to be \emph{locally compatible}
with an LC structure $p\mapsto\sC_p$, iff, with $\sC_p \supset \phi_p(N)\cong N$, it holds
\beq{chardir}
g_p(V(p),V(p))=0\Leftrightarrow V(p)\in T_p\phi_p(N),\ \forall p\in\Int M  ,
\eeq
i.e. locally at any vertex the cone determines
the characteristic null directions
in the tangent space.

On the other hand, the cone structure poses an equivalence relation on
Lorentzian metrics which are compatible with the LC structure.
Given any such metric $g$,
the corresponding equivalence class $[g]$ is
the conformal class of $g$.
We summarize the existence and uniqueness result as follows:
\\
\textbf{Proposition 2:}
Given a strong LC structure on a (CAT) manifold,
\\
(i) there always exist a (CAT$_{-1}$) Lorentzian metric $g$ on $M$
compatible with the LC structure.
\\
(ii) the conformal class $[g]$ of LC compatible metrics is uniquely
determined by the LC structure.

The {existence}
of a conformal Lorentzian metric
is guaranteed by a {\em strong} LC structure,
but not by a weak one.
However,
since conditions \ref{charsurf} needs only the existence of the tangent bundle
of $\sC_p$,
{uniqueness} is assured already by a differentiable \emph{weak} LC
structure.

Although
at each $p\in\Int M$
a CAT strong LC structure on $M$ admits
a conformal class $[g]$ of CAT$_{-1}$ Lorentzian metrics $g$ with
characteristic directions in $T_p M$ tangential to $\sC_p$,
away from the vertex $p$ the cones of the LC structure
need not at all be compatible with the null structure of any
conformal metric $[g]$.
This reflect the fact that, apart from its local vertex structure,
a strong LC structure is still much more flexible than a conformal structure.
For any $q\neq p$
the tangent directions given by $T_q\sC_p$ need a priori
not be related
to tangent directions of null curves of $g$,
since the cone (or its thickening) at $p$ is
in general unrelated to that at $q$.
The need for  compatibility conditions
between cones at different points
motivates
the introduction of some of the
causality structures in open regions of $M$
introduced later in the following section.
\section{Causality structures on manifolds}
\setcounter{equation}{0}
Given a (weak or strong) LC structure one wants to know first of all
under which conditions, for given $p\in\Int  M$
an exterior and interior of the cone can be distinguished
within the complement $M-\sC_p$.
This problem is the global analogue of the local
one which was answered by Proposition 1 and Theorem 1 above.

\textbf{Proposition 3:}
Assume that at  $p\in\Int  M$ there are open (CAT) submanifolds
$\sT_p$ and $\sE_p$ such that
the interior of $M$ decomposes into the disjoint union
$\Int M =\sC_p \dcup \sT_p \dcup \sE_p$.
Assume $\exists$ $k\in\N_0: \Pi_{k}(\sT_p)\neq \Pi_{k}(\sE_p)$.
Then $\sT_p$ and $\sE_p$ can be topologically distinguished.
\\
\emph{Proof:}
$\Int M-\sC_p=\sT_p \dcup \sE_p$, and homotopy groups are topological
invariants.
\hfill \mbox{$\square$}

In particular the conditions of Proposition 3 apply
for $d+1>2$ in particular to all manifolds with the
following topological structure:

{\bf Example 1:}
Let in any dimension $d:=\dim M-1>0$
at any $p\in\Int  M$ be
$|\Pi_{0}(\sT_p)|=2$ and
$\Pi_{d-1}(\sE_p)=\Pi_{d-1}(S^{d-1})$,
for $d>1$ be $\Pi_{d-1}(\sT_p)=0$ and $|\Pi_{0}(\sE_p)|=1$,
For $d=1$ be $\Pi_{d-1}(\sT_p)=\Pi_{d-1}(\sE_p)=\Pi_{0}(S^{0})$,
i.e. $|\Pi_{0}(\sT_p)|=|\Pi_{0}(\sE_p)|=2$, and
in dimension $d=0$ be $\sT_p=\sE_p=\emptyset$ at any $p\in\Int  M$.
Then in particular $\sT_p\not\cong\sE_p$ $\iff$ $d+1>2$.
The open exterior $\sE_p$ and
the open interior $\sT_p$ of the cone $\sC_p$
at any interior point $p\in\Int  M$
are topologically distinct for $d>1$,
indistinguishable
for $d=1$, and empty for $d=0$.

In the case of Example 1,
$\sC_p$ divides $M\setminus \sC_p$ in three
open submanifolds,
a non-contractable exterior $\sE_p$,
plus two contractable connected components of
$\sT_p=:\sF_p\cup\sP_p$, the future
$\sF_p$ and the past $\sP_p$ with
$\partl \sF_p=\sC^+_p:=\phi_p \sC^+$
and $\partl \sP_p=\sC^-_p:=\phi_p \sC^-$
respectively.
This rises also the question if and how $\sF_p$
and $\sP_p$ can be distinguished.

Let $M$ be differentiable
and $\tau$ be any vector field  $M\to TM$
such that at any $p\in\Int  M$ its orientation
agrees with that of $\phi_p(a)$.
Such a orientation vector field comes naturally along with a
(CAT$_{-1}$) $\Z_2$-connection on $M$
which allows to compare the orientation $\tau(p)$ at different $p\in\Int  M$.
Given a strong LC structure on $M$,
the $\Z_2$-connection is in fact provided via continuity of $p\mapsto T_p\phi_p(a)$.
Then $\tau$ is tangent to an integral curve segment
through $p$ from  $\sP_p$ to $\sF_p$.
In particular,
$\sF_p$ and $\sP_p$ are distinguished from
each other by a consistent $\tau$-orientation on $M$.

If $M$ is not differentiable,
in order to distinguish continuously $\sP_p$ from $\sF_p$ on $\Int M$
it remains just to impose
a topological  $\Z_2$-connection on $\Int M$ a fortiori.

In order to obtain a more specific causal structure
it remains to identify natural consistency conditions
for the intersections of cones of different points.
In order to define topologically
timelike, lightlike, and spacelike relations,
and a reasonable causal complement,
we introduce the following causal consistency
conditions on cones.
\\
\textbf{ Definition 8:}
$M$
is \emph{(locally) cone causal} or \emph{C-causal} in an open region $U$,
if it 
carries a (weak or strong) LC structure
and in  $U$
the following relations between different cones in $\Int M$ hold:
\\
(1) For $p\neq q$ one and only one of the following is true:
\\
(i) $q$ and $p$ are causally \emph{timelike} related, $q\ll p$ $:\Iff$
$ q\in \sF_p$ $\wedge$ $p \in \sP_q $
(or $p \ll  q$)
\\
(ii) $q$ and $p$ are causally \emph{lightlike} related, $q\lhd  p$  $:\Iff$
$ q\in \sC^+_p\setminus \{p\}$ $\wedge$ $p \in \sC^-_q\setminus \{q\} $
(or $p \lhd  q$),
\\
(iii) $q$ and $p$ are causally unrelated,
i.e. relatively \emph{spacelike} to each other, $q\bowtie  p$ $:\Iff$
$ q\in \sE_p$ $\wedge$ $p \in \sE_q $.
\\
(2) Other cases (in particular non symmetric ones)
do not occur.
\\
$M$ is {\it locally C-causal}, if it is C-causal in any  region $U\subset M$.
$M$ is {\it C-causal} if conditions (1) and (2) hold $\forall p\in \sC$.

Let $M$ be C-causal in $U$.
Then, $q \ll  p$ $\Iff$ $\exists r: q\in \sP_r \wedge p\in \sF_r $,
and $q \lhd  p$ $\Iff$ $\exists r: q\in \sC^+_r \wedge p\in \sC^-_r $.

If an open curve $\R\ni s\mapsto c(s)$ or a
closed curve $S^1\ni s\mapsto c(s)$ is embedded in $M$, then in particular
its image is $\Img(c)\equiv c(\R)\cong \R$ or
$\Img(c)\equiv c(S^1)\cong S^1$ respectively,
whence it is free of self-intersections.
Such a curve is called \emph{spacelike}
$:\Iff$ $\forall p\equiv c(s)\in \Img(c) \exists \epsilon :
c|_{]s-\epsilon,s+\epsilon[\setminus\{s\}}\in\sE_{c(s)}$,
and \emph{timelike}
$:\Iff$ $\forall p\equiv c(s)\in \Img(c) \exists \epsilon :
c|_{]s-\epsilon,s+\epsilon[\setminus\{s\}}\in\sT_{c(s)}$.

Note that C-causality of $M$ forbids a multiple refolding intersection
topology for any two cones.
In particular along any timelike curve
the future/past cones do not intersect,
because otherwise there would exist points which are simultaneously
timelike and lightlike related.
Continuity then implies that future/past cones in fact foliate the part of $M$
which they cover. Hence, if there exists  a fibration
$\R\hookrightarrow\Int M\twoheadrightarrow \Sigma$,
then C-causality implies that the future/past cones foliate in particular
on any fiber. In fact, given a fibration,
one could define also a weaker form of causality
just by the foliating property of all future/past cones on each fiber.
(Physically this situation corresponds to ultralocal
classical clocks. Quantum uncertainty of the fiber
would require to take appropriate ensemble averages over
some  bundle of neighboring fibers which then contains in particular
spacelike related vertices on the fibers of the bundle. Then
the corresponding future or past cones intersect for sure,
and even timelike related ones of different fibers \emph{may} intersect !)
C-causality however requires more, namely the future/past cones of {\em all}  timelike
related vertices should be non-intersecting, not only those in a particular fiber.

Therefore C-causality
allows also a reasonable definition of  a causal complement.
\\
\textbf{Definition 9:}
For any open set $S$ in a C-causal manifold $M$
the \emph{causal complement}
is defined as
\beq{cc}
S^\perp:=\bigcap_{p\in \cl S}{ \sE_p} ,
\eeq
where $\cl S$ denotes the closure in the topology
induced from the manifold.
Although the causal complement is always  open,
it will in general  not be a contractable region even if $S$ itself
is so.

Assume $p$ and $q$ are timelike related, $p\in\sP_q$ and $q\in \sF_p$.
$\sK^q_p:=\sF_p\cap\sP_q$ is the bounded open double cone between $p$ and $q$.
Given any bounded open $\sK$ such that $\exists p,q\in M: \sK=\sF_p\cap\sP_q$,
we set $i^+(\sK):=\{q\}$,  $i^-(\sK):=\{p\}$, and $i^0(\sK):=\sC^+_p\cap\sC^-_q$.
For any $\sK^q_p\subset M$ let
$\clc(\sK^q_p)
$
be its {\it causal closure}.

Since C-causality prohibits relative refolding of cones, it also ensures that
$(\sK^q_p)^{\perp\perp}=\sK^q_p$, i.e. the causal complement is a duality
operation on double cones.

The open double cones of a C-causal manifold $M$ generate a topology,
called the {\it double cone topology}
which is a genuine generalization of the usual Alexandrov topology
for strongly causal space-times.
For strongly causal space-times the Alexandrov topology is equivalent
to the manifold topology
\cite{KP,Pen72}.
When $M$ fails to be locally causal
the double cone topology may be coarser than the manifold topology.

Let us discuss now possible natural relations
that can appear between two double cones $\sK_1$ and $\sK_2$
of a C-causal manifold.
First there is the case
$\sK_{1}\cap \sK_{2} =\emptyset$
which corresponds to  causally unrelated
sets.
For $\sK_{1}\cap \sK_{2} \neq\emptyset$,
the intersection is such that
$
\sK_{1}\cup \sK_{2} \setminus \sK_{1}\cap \sK_{2}
$
is either given by two disconnected pieces
or it is connected.
In the latter case we distinguish
whether
$
\partl \sK_{1}\cap \partl \sK_{2}
$
is empty or not.
It is in the former case that
one of $\sK_1$ and $\sK_2$ will be contained in the other.

Local C-causality does a priori not preclude
other more pathological possibilities.
However it is possible to define in a purely topological manner
more refined causality notions.

\textbf{Definition 10:} Let $M$ be a C-causal manifold.
\nl
(i)
$M$ is {\it globally hyperbolic} $:\Iff$ $\clc{\sK^q_p}$ compact $\forall p,q\in M$
\nl
(ii)
$M$ is {\it causally simple} $:\Iff$ $\clc{\sK^q_p}$ closed $\forall p,q\in M$
\nl
(iii)
$M$ is {\it causally continuous} $:\Iff$ $M$ is distinguishing and both
$\sF: p\mapsto \sF_p$ and $\sP: q\mapsto \sP_q$
are continuous
\nl
(iv)
$M$ is {\it stably causal}  $:\Iff$ $M$ admits a $\CO$ function $f:M\to\R$
strictly monotoneously increasing along each future
directed nonspacelike curve (global time function)
\nl
(v)
$M$ is {\it strongly causal} $:\Iff$  the topology generated by $\{\sK^q_p\}_{p,q\in M}$
is equivalent to the manifold topology of $M$
\nl
(vi)
$M$ is {\it distinguishing} $:\Iff$
$\sF_p=\sF_q \Imp p=q  \wedge  \sP_p=\sP_q \Imp p=q$
\nl
(vii)
$M$ is {\it causal} $:\Iff$ every closed curve in $M$ is not nonspacelike
\nl
(viii)
$M$ is {\it chronological} $:\Iff$ every closed curve  in $M$ is not timelike

If a manifold carries a Lorentzian metric, we saw in Section II
above that this is locally compatible with a strong LC structure.
Beyond that,
it is an interesting question under which conditions a
Lorentzian metric is \emph{compatible} with some LC structure.
The Minkowski metric $\eta$ is  {compatible} with the cone $\sC$ in the sense that
$\eta_x(v,v)=0$
$\Leftrightarrow$
$(x,v)\in T\sC:=\bigcup_{y\in \sC}{T_y \sC}$
where ${T_y \sC}:=\bigcup_{y\in N\subset\sC}{T_y N}\subset\R^{d+1}$
and the latter union is over all (differentiable) $1$-dimensional submanifolds $N\subset\sC$
passing through $y$,
with  all their tangent spaces embedded as linear submanifolds with common origin within
the common embedding space $\R^{d+1}$.
Hence, for $y\neq 0$, the fibers $T_y \sC\cong\R^{d}$ are all usual isomorphic tangent spaces,
while the only non-standard fiber $T_0 \sC\cong \sC\subset\R^{d+1}$ reproduces the $d$-dimensional cone itself,
which is the  local model of its own singularity.
Correspondingly, a Lorentzian metric $g$ is said to be \emph{compatible} with some LC structure
$p\mapsto\sC_p$, iff
\beq{charsurf}g_q(V(q),V(q))=0 \
\Leftrightarrow \
\left[
\forall p\in M:
 q\in\sC_p
\Rightarrow
V(q)\in T_q\sC_p :={(\phi_p)}_{*}T_{{\phi_p}^{-1}(q)} \sC=\bigcup_{{\phi_p}^{-1}(q)\in N\subset\sC}{T_{q}\phi_p (N) }
\right]\  ,
\eeq
where the latter union is over all (differentiable) $1$-dimensional submanifolds $N\subset\sC$
passing through ${\phi_p}^{-1}(q)$,
and the latter identity holds with tangent push forward $(\phi_p)_{*}T_y N:=T_{\phi_p(y)}\phi_p (N)$.
Therefore, with \rf{charsurf} the cones are the characteristic surfaces of the
Lorentzian metric.
As pointed out above, \rf{charsurf} does not hold in
general.
However one might search for sufficient and necessary
causality conditions such that this compatibility holds. A
systematic investigation of this point is beyond our present
investigations. Let us rather assure
the correspondence of the causality notions
of Def. 10 to the usual ones in the case of a Lorentzian space-time.

\textbf{Theorem 2:} Let $M$
carry a smooth Lorentzian metric $g$.
Then the Lorentzian metric determines a C-causal structure.
If a C-causal structure of $M$ is related to a Lorentz metric,
then the definitions (i)-(viii) agree with the standard definitions
and the following chain of implications of properties of $M$ holds:
globally hyperbolic $\Imp$
causally simple $\Imp$
causally continuous $\Imp$
stably causal $\Imp$
strongly causal $\Imp$
distinguishing $\Imp$
causal $\Imp$
chronological.

{\it Proof:}
Given a smooth Lorentzian metric $g$
the cones determined by the null structure $[g]$ respect
the relations of Def. 8, because otherwise
there would exist some singularities.
For (v) in the case of Lorentzian manifolds see  \cite{Pen72},
for the other notions
and for the chain of implications
see \cite{BEE}.
\hfill \mbox{$\square$}

Finally let us define a condition which excludes the existence of
singularities or internal boundaries within the future and past cones.

\textbf{Definition 11:}
Let $M$ carry a (weak or strong) LC structure.
\\
(i)
$M$ is {\it precausal} in an open region $U\subset M$,
if the $d+1$-parameter CAT
family $\{\phi_p\}_{p\in U}$
of CAT homeomorphisms  $\phi_p:\R^{d+1}\supset V\to U$
is such that at any $p\in U$ it is $\sC_p|_U=\phi_p \sC|_V$,
and any  CAT submanifold of $\sC_p$ or $(M-\sC_p)\cap U$
is a CAT homeomorphic image of  $\sC$ or $(\R^{d+1}-\sC)\cap V$  respectively.
$M$ is {\it locally precausal} iff it is precausal in any open region $U\subset M$.
\\
(ii)
$M$ is {\it precausal}
if
it is locally precausal
such that
in the CAT $d+1$-parameter family $\{\phi_p\}_{p\in U}$
any CAT homeomorphism extends also to a homeomorphism
of the interior
$\phi_p: \sE\to \sE_p$.



\section{Discussion and perspective} 
\setcounter{equation}{0}
Above we presented topological definitions of
local (i.e. pointwise) cone (LC) structures
for a general topological or differentiable manifold $M$
of dimension $d+1>2$ and notions of { causality} on $M$
in a purely topological manner.
It is remarkable that such definitions are possible, whence  the usual
recursion to a Lorentzian metric becomes redundant.

Proposition 1 gives criteria which \emph{locally} distinguish the exterior
and the interior of the cone at any point from each other.
Proposition 3 and Example 1
provide concrete global topological conditions for $M$
in order to allow the relative distinction of interior and exterior of all its
cones.
Minkowski space is obviously a manifold which satisfies the
conditions for topologically distinguished interior and exterior
according to  Example 1. It is however a priori not clear what
for each given category CAT of manifolds
is the largest class of manifolds with
the topological structure described in Example 1.

We saw that a global consistent distinction between future and past cones
requires just a topological $\Z_2$-connection.
Note that, as an important possible application,
the canonical approach to quantum gravity
comes always along with such a connection.
In fact the canonical configuration variables for oriented manifolds may be
there be chosen as $\SO(d+1)$-connections.

The presented LC  structures,
C-causality, and other our purely topological causality
notions provide some alternative to the poset approach
\cite{BLMS,MS,Mar99,RS}
for defining causality in
quantum theories of quantum gravity.
While that approach is based on a much weaker local notion of causality
on \emph{sets},
which essentially involves only a partial ordering,
the present definition gives the possibility to work with
local definition of causality on differentiable \emph{manifolds}
which still captures the essential notions for
curves in a C-causal manifold
to be lightlike, timelike or spacelike
without the need of an underlying  Lorentzian structure.
For any set $S$ in a C-causal manifold
a topological notion of a causal complement $S^\perp$
is given by (3.1).
Any double cone $\sK$ in a C-causal manifold then has the
duality property
$\sK^{\perp\perp}=\sK$.

Some advantages of conformal invariance in the quantization in
minisuperspace models of higher-dimensional Einstein gravity have
been pointed out in \cite{Ra1,Ra2}. In particular, factor ordering
problems can be resolved uniquely this way. For a more general
background independent quantum theory the restriction of local
diffeomorphisms to those consistent with a causal structure, say
e.g. a LC structure, on the whole manifold might appear too
restrictive. After all a strong LC structure implies already the
existence of a conformal metric, whence diffeomorphisms may be
restricted locally to conformal ones. Nevertheless note that even
a strong LC structure is much more flexible than a conformal
metric structure. The local cones of different vertices might
refold away from their vertices with rather complicated
intersection topologies while a CAT continuous conformal metric
within its (regular!) domain
does not admit refolding singularities of the characteristic
surfaces, each of the which is spanned out by all the null geodesics passing
through a given vertex.
Of course refolding and the associated singularities should be
a topic of further more systematic investigations elsewhere.

The canonical approach to field quantization usually employs a
foliation $\Sigma\hookrightarrow\Int M\twoheadrightarrow \R$.
This rises the question when this is consistent
with a (C-)causal structure. This may roughly be answered as follows:
A CAT foliation of $M$ may be said to be consistent with a C-causal structure,
if for any open set $O$ in a CAT slice $\Sigma\subset M$
there exists
a double cone $\sK\subset \Int M$                      
such that $i^0(\sK)\subset \partial(\sK\cap\Sigma)$
(compare also Section III, below Def. 9).

Consider now such a double cone $\sK$ in $M$
with $O:=\sK\cap\Sigma$ and $\partl O=i^0(\sK)$
and a diffeomorphism $\phi$ in $M$ with pullbacks
$\phi^\Sigma\in\Diff(\Sigma)$ to $\Sigma$
and
$\phi^{\sK}\in\Diff(\Sigma)$ to $\sK$.
If $\phi(\sK)=\sK$, it can be naturally identified
with an element of $\Diff(\sK)$.
($\phi|_{M\setminus \sK}=id_{M\setminus \sK}$ is a sufficient
but not necessary condition for that
to be true.)
If in addition  $\phi(\Sigma)=\Sigma$
then also $\phi(O)=O$, and $\phi|_O$ is a diffeomorphism of $O$.

When $M$ is homeomorphic to $\Sigma\times \R$
it is straightforward to extend
the above from a single
hypersurface $\Sigma\subset M$ to a foliation of $M$
via a $1$-parameter set of embeddings $\Sigma\hookrightarrow M$.

For a canonical approach to quantum gravity,
one might want to work with a restriction
of the causal structure to cones with their
vertices on a given graph $\Gamma$
within a slice $\Sigma$ of a foliation.
A given  topological (differentiable) causal structure,
selects particular causal homeomorphisms (diffeomorphisms)
which preserve it.
A strong LC-structure on all of $M$ already implies
the existence of a conformal metric structure
and a requirement of compatibility with that metric
would reduce the local covariance group to local conformal
diffeomorphisms. One might however also weaken the LC and causal structure
of the manifold by considering  in any leaf $\Sigma$ of a given foliation
only cones with vertices on $\Gamma\subset\Sigma$
instead on all of $\Sigma$.
A natural choice for $\Gamma$ is the dual graph
of a triangulation.
Then the cones have to CAT vary  along the edges,
but at least for CAT$\supset \Cinf$
the cones at the vertices of the graph can be freely ascribed.
Consequently, a geometry constructed on that basis
will be invariant under diffeomorphisms much more general
than conformal ones.

Let us however also emphasize that,
although the existence of a local conformal metric is guaranteed by a
strong LC structure,
it is a priori not obvious that this metric should play any significant
r\^{o}le. Then however also the need to restrict diffeomorphisms to those
compatible with the conformal metric may be questioned.
One might eventually expect that
within some approach to quantum geometry
a cone at a vertex $p\in O\subset\Sigma$
should be replaced by an appropriate average
over cones with vertices within some  region $O$ of minimal
Heisenberg uncertainty. Then the flexibility of the weak and strong
LC structures makes them  interesting concepts
and potential ingredients for a possible definition of
quantum causality too. Presently however this
is still matter of many speculations.

Classically,
the existence of a local metric requires only the
differentiable structure in
an arbitrary small neighborhood of the vertex,
and the defined  LC structures
fix the preferred null directions
only locally at each vertex.
With sufficiently strong notions of causality
(e.g. C-causality above) the null structures of this
metric may become  consistent
with the global structure of cones of the LC structure.
Note that in the case of a given Lorentz metric
null geodesics lie on cones,
and with sufficiently strong causality, e.g. global hyperbolicity,
these cones have to be consistent with respect to each other
and under variation of the vertex without refolding into each other
(i.e. in particular without conjugate points).

For Lorentzian manifolds there is a hierarchy of common notions of causality
which have been generalized above.
Provided our definitions of causality are  sufficiently natural
it should be possible to prove (at least parts) of this hierarchy
in the more general topological setting. However a complete
investigation of the mutual relations between
different topological causality concepts
is beyond the scope and goal of the present
paper.

It should be emphasized that the above was just
brief demonstration of the possibility to introduce
notions of cones and causality on CAT topological  manifolds
without a metric. In particular,
weak and strong LC structures, C-causality, precausality,
and some generalizations of the most common notions of causality
have been obtained.
However the investigation is far from complete.
It remains for future work to develop the topological approach
to causal structures on manifolds further, to investigate better
some of its implications,
and last not least to demonstrate its
applicability in background independent formulations of
algebraic quantum field theory and quantum gravity.

\section*{Acknowledgements}
I wish to thank to O. Dreyer, M. Holman,
R. Puzio, and M. Reisenberger for discussions, and to N. Guerras for continuous
support.

\nl
\nl

\end{document}